\documentclass[twocolumn,aps,amsmath,amssymb]{revtex4}

\newlength\picturewidth
\usepackage[latin1]{inputenc} \usepackage{amsmath} \usepackage{amssymb}
\usepackage{amscd} \usepackage{graphicx} \usepackage{subfigure}
\usepackage{epstopdf} \usepackage{graphicx}
\usepackage{color}

\begin{document}

\title{Phase behavior of repulsive polymer-tethered colloids }

\author{Behnaz Bozorgui, Maya Sen, William L. Miller, Josep C. P\`{a}mies and Angelo Cacciuto} 
\email{ac2822@columbia.edu}
\affiliation{Department of Chemistry, Columbia University, 3000 Broadway, \\New York, New York 10027}
\date{\today}

\begin{abstract}

We report molecular dynamics simulations of a system of repulsive, polymer-tethered colloidal particles.
We use an explicit polymer model to explore how the length 
and the behavior of the polymer (ideal or self-avoiding) affect 
the ability of the particles to organize into ordered structures when the system is compressed to  
moderate volume fractions. We find a variety of different phases whose 
origin can be explained in terms of the configurational entropy of polymers and
colloids. Finally,  we discuss  and compare our results to those obtained for similar systems using 
simplified coarse-grained polymer models, and set the limits of their applicability.
\end{abstract}
\maketitle

\section{Introduction}

Understanding how colloidal particles spontaneously organize into ordered macroscopic aggregates
is a longstanding challenge that has recently acquired an extra degree of complexity.
In fact,  advances in particle synthesis~\cite{DeVries,Schnablegger,Hong,Weller,Hobbie} have opened the way to the production of 
colloidal particles that are anisotropic both in shape and surface chemistry.  
This provides an unlimited number of building blocks that can spontaneously assemble 
into an unprecedented variety of structures with potentially  novel functional, 
mechanical, and optical properties.

The effect of the anisotropy of nanoparticles on their macroscopic ordering can be addressed in terms of 
(a) the form of the inter-particle interaction, and (b) their shape. Not surprisingly, for both cases there is ample  evidence (see for example ~\cite{glotzer,chandler,geissler,cacciuto,torquato,frenkel,glotzer2,glotzer3,glotzer4,esco} and references therein) 
of a strong correlation between the physical properties of the components and those of the resulting aggregates.
This phenomenology must be thoroughly explored as it may lead the way to a rational 
design of the components to target desired macroscopic structures.
 
Here we focus on the role of particle shape. Specifically, 
we study the phase behavior of a particularly interesting class of deformable particles that 
is obtained by grafting a {single} long chain to a {each} colloid. What makes this hybrid colloid
intriguing is that, because of the flexibility of the polymer, the overall shape of 
the particle is not fixed, but can be
spontaneously altered depending on the specific thermodynamic states imposed on the system. 
{A few experimental realizations of particles compatible with our model have recently been 
synthesized~\cite{exp1,exp2,exp3}.} 
The dual nature of these nanoparticles may open the door to exotic self-assembled structures 
that are not typically seen in systems of nanoparticles with intrinsic (invariable) shape, {and its 
key elements are very similar to those of asymmetric diblock copolymers.}  

Unlike recent experimental  and theoretical  studies on particles coated with dsDNA, 
which can form complex networks between the particles 
via linker-mediated dsDNA-dsDNA interactions\cite{Nykypanchuk,Park,Bozorgui}, 
no explicit attractive forces are introduced in our system. 
As a  result, any phase described in this paper will be mostly driven by a nontrivial
balance between the configurational entropy of the colloids and that of the chains.

In this paper we use molecular dynamics simulations to understand 
the phase behavior of a system of repulsive, polymer-tethered colloidal particles. 
Specifically, we consider a system in which 
each colloid is connected to one of the end groups of a single polymer, and we study how different structures emerge depending on the 
polymer length. Furthermore, we explicitly analyze both the case of ideal and self-avoiding polymers.

Capone {\it et al.}~\cite{Capone} have recently analyzed the phase behavior of model of di-block copolymers. 
Their study bears similarities with our work, as some of the self-assembled structures are 
common to both systems. However, the two systems differ in the way the polymers are modeled. 
Here we use an explicit beads-and-springs model, while spheres with a soft potential were employed 
by Capone {\it et al}.
Though computationally expensive, our model enables us to gain a detailed understanding of 
the mechanisms behind the nontrivial phase behavior emerging in this system, and  this choice 
will turn out to be quite critical when considering 
the case of non-ideal polymers.
To the best of our knowledge this is the first computational study that explicitly accounts 
for the internal degrees of freedom of the polymer for this particular system.

\section{Model}
\label{sec:sim}
\begin{figure}
  \centering \includegraphics[width=\picturewidth]{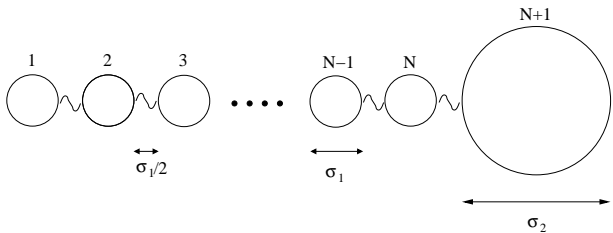}
  \caption{Schematic representation of our model for a hybrid, polymer-grafted colloid. 
  The first $N$ monomers of diameter $\sigma_1$ represent the chain, whereas the $N+1^{\text{st}}$ monomer 
of diameter $\sigma_2$ represents the colloidal particle. Both particles are assumed to have equal mass. }
  \label{fig:1}
\end{figure}

We model the polymer-tethered colloids as a polymer of $N+1$ monomers, 
with $N$ monomers of diameter $\sigma_1$, and monomer $N+1$, representing the colloidal particle, 
of diameter $\sigma_2$.
See Fig.~\ref{fig:1} for a depiction of the particle. 
In this model the $N^{th}$ monomer is not constrained to be at a specific location on the surface of the colloid, 
but can freely diffuse on it; constraining this monomer would yield the same equilibrium properties.

Excluded volume interactions between any two particles  in the system are enforced via a purely repulsive
shifted-truncated Lennard-Jones potential

\begin{equation}
  U^{\text{E}}_{i,j}(r)=\epsilon_{i,j}\left[\left(\frac{\sigma_{ {i,j}}}{r_{i,j}}\right)^{12} -
    \left(\frac{\sigma_{{i,j}}}{r_{i,j}}\right)^6 + \frac{1}{4}\right]  \, ,
\hspace{.5cm} \forall\,\, \, r_{i,j}\leq 2^{1/6}\sigma_{i,j}
  \label{eq:WCA}
\end{equation}

The indices $i,j\in\{1,2\}$ indicate the identity of the particle (polymer or colloid, respectively.) 
$\sigma_{i,j}\equiv (\sigma_1+\sigma_2)/2$. $r_{i,j}$ is the distance between the centers of mass of any two particles. 
Finally, $\epsilon_{i,j}=10 k_{\text B}T$ $\forall$  $i,j$ when
considering self-avoiding polymers, and we set  $\epsilon_{1,1}=0 k_{\text B}T$ for the case of ideal polymers. 

In each hybrid colloid, particles are linearly connected via the harmonic spring potential 
\begin{equation}
U^{\text{S}}_{i,i+1}=k_s(r_{i,i+1}-r^0_{i,i+1} )^2
\end{equation}
where $k_s=150 k_{\text B}T$ is the spring constant, and $r^0_{i,i+1}=\sigma_{i,i+1}+\sigma_1/2$ is the equilibrium distance.

We perform $NPT$ molecular dynamics simulations using the LAMMPS (Large-scale Atomic/Molecular Massively Parallel Simulator) package~\cite{lammps}.
Pressure and temperature are kept constant by means of a Nos\'{e}-Hoover thermostat~\cite{NH1} and
barostat~\cite{NH2} with  additional drag terms, with coefficients 
$\xi_T=1 \tau_0^{-1}$ and $\xi_P=1 \tau_0^{-1}$ respectively ($\tau_0$ is the reduced time unit), 
to damp the dynamics and suppress large temperature and pressure oscillations.

The simulation box is a cuboid with periodic boundary conditions and, for  pressure control, we use decoupled box lengths in each of the 
three Cartesian coordinates. This allows box aspect ratios to vary slightly to accommodate crystalline structures.

The system initial configurations are prepared by performing $NVT$ simulations in the gas phase.
Once the system is equilibrated, and the initial pressure $P_0$ is extracted from the thermalized configurations, 
we slowly ramp  the pressure to the desired value $P_1$ starting from $P_0$  (all pressures referred in this paper are rescaled with respect to 
the colloidal interaction energy $\varepsilon_{22}$ and the colloidal diameter $\sigma_2$).
Each  subsequent simulation performed at a constant pressure $P_i$ starts from the thermalized configuration 
at pressure $P_{i-1}$ ($P_{i-1}<P_{i}$). {The dimensionless $\Delta P$ was typically set to 0.01, and was refined or extended depending on the 
distance from the transition point.}
This procedure ensures that the chains have the time to fully equilibrate.  
{The relatively short chain lengths considered in this study, and the relative low densities at which most of the phases occur, 
result in a system that does not have a pathological dynamical behavior. 
Therefore, the very slow pressure annealing described above, together with the monitoring of all the observables considered in this study, 
including the average cluster size and its distribution where sufficient to establish equilibrium. 
Moreover, we checked that we could reproduce all the phases starting from completely different initial configurations.}

In our study we considered tethers with a minimum of $N=5$ and a maximum of 
$N=300$ monomers, and colloids of diameter ranging from $\sigma_2=2\sigma_1$ to $\sigma_2=18\sigma_1$.

All of our simulations are carried out using a total of 512 hybrid colloids at room temperature, and the 
longest simulations took about six months of computer time on an Intel Xeon X5355 2.66GHz processor. {In all simulations we set the time step to
 $\Delta t = 0.015 \tau_0$.  For each pressure annealing step, $\Delta P$, simulations were run for a minimum of $10^6$ (for the small chains) to a maximum of $10^8$ timesteps 
(for the long chains.)}

Every observable reported in this paper is expressed in dimensionless units.


\section{Results}

Apart from the harmonic potential, which serves a purely structural purpose by enforcing connectivity between the different components of our hybrid colloid,
 there are no attractive interactions in our system.
As a consequence, the free energy is dominated, at the low concentrations considered in our study, by the configurational entropy of its components.
Although the configurational entropy is, strictly speaking, associated with hard potentials,
 we have chosen a large value for $k_s$ to ensure that  bonds are very close to their equilibrium length, and 
have also run a few simulations with a stricter excluded volume constraint by setting 
$\epsilon_{i,j}=500 k_{\text B}T$. We find no discernible difference between the two cases under several thermodynamic conditions.

What follows are the  phase diagrams for ideal and self-avoiding tethers as a function of the volume fraction of colloids, $\phi\equiv \pi\sigma_2^3 N_c/(6V)$, and the effective polymer-colloid size ratio, 
which we define as $\alpha = 2 R_g / \sigma_2$. $N_c = 512$ is the number of colloids, $V$ the volume of the simulation box, 
and  $R_g$ is the radius of gyration of a polymer tether, which scales as $R_g \sim (N)^{1/2}$ for ideal polymers and as $R_g\sim N^{3/5}$ for self-avoiding ones. 
{The calculated bulk prefactors  for our models  are 0.57 and 0.60 for  ideal and self-avoiding chains respectively.} 
{Experimentally, one can easily control the behavior of the polymer by altering 
the properties of the solvent. For instance, the polymer will behave ideally at the solvent $\theta$ point, and as a SAW at larger temperatures.}

\subsection*{Ideal chains}
\begin{figure*}
    \includegraphics[width=2.0\picturewidth]{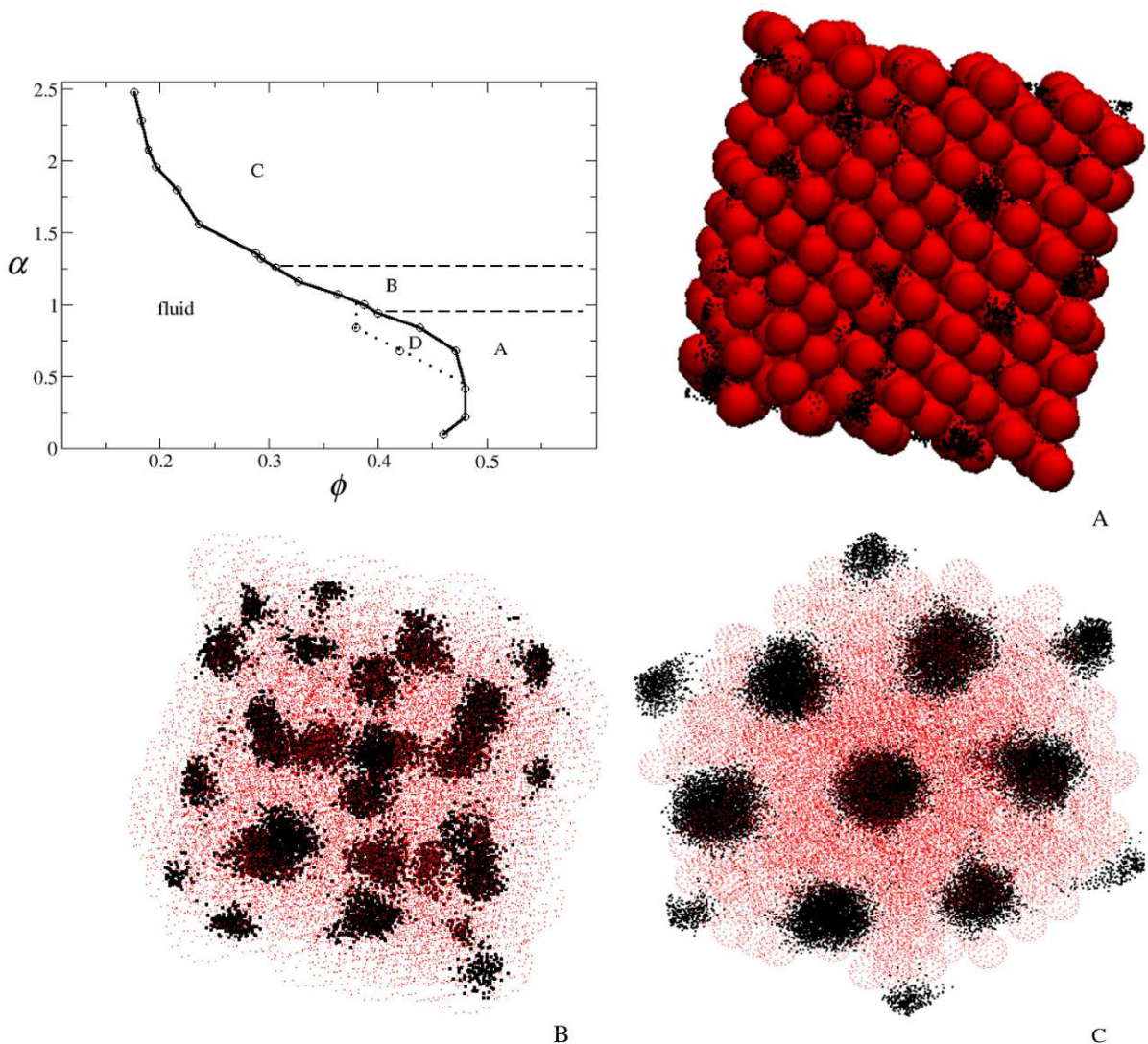}
    \caption{(Color online)  Phase diagram of colloids with ideal tethers as a function of the polymer-colloid size ratio $\alpha$ and colloid volume fraction $\phi$. 
 Snapshots of the phases in region (A), (B), and (C), depicting 
the colloidal crystal, the disordered micellar, and the micellar crystal phase, respectively, are also shown. 
For the sake of clarity, in snapshot (B) and (C), the colloidal particles are depicted using a light, low-density pixel representation, while 
the dark regions show where the polymer chains are located.
 }
\label{fig:snap1}   
\end{figure*}

Figure \ref{fig:snap1},
obtained using several combinations
of colloidal radii and chain lengths, shows the different phases arising from the organization of the particles in the system
as a function of volume fraction for different values of $\alpha$, and presents several interesting features.
{The  lines are a guide to the eye, and are found by identifying the threshold densities and size ratios above which phase change occurred.}

For $\alpha$ sufficiently small, $\alpha \lesssim 1$, the presence of the tethers does not alter 
the ability of the colloids to crystallize into a macroscopic FCC crystal once the system is compressed 
above a threshold volume fraction. This is exactly how tether-free colloids crystallize under analogous 
conditions, and is achieved in our system by chain localization into either the interstitial
space between the colloids (for very small $\alpha$) or into crystal vacancies as depicted
in Fig.~\ref{fig:snap1}A. This is only possible
 as long as the  chains are short enough to fit within a vacancy without 
exerting a significant amount of pressure arising from chain confinement.
The formation of crystal vacancies is the first hint of colloidal/polymer segregation. 
This phase is preceded by a fluid phase of small micelles at a lower volume fraction
(Fig.~\ref{fig:snap1} region D). These deform and freeze as the system pressure is increased into structurally FCC-compatible cages: 
the vacancies in the colloidal crystal lattice.
Each vacancy is typically filled by the polymer chains of all colloids surrounding it, and their locations present no obvious 
translational order. In fact, we find a non-negligible number of vacancy pairs distributed across the colloidal crystal.

Interestingly, for $1 \lesssim\alpha\lesssim 1.3$, the colloidal crystal phase ceases to form, and is replaced by 
a disordered micellar phase (see Fig.~\ref{fig:snap1}B). This is clearly due to the increased free energy cost 
associated with chain confinement into a vacancy which grows quadratically 
with $\alpha$,\cite{deGennes} $\Delta F\sim n(2R_{\text g}/\sigma_2)^2=n\alpha^2$, 
where $n$ is the number of chains in the same vacancy. To mitigate this effect, 
the typical cage sizes become larger and the geometrical rearrangement into an FCC-cell becomes
 expensive. The presence of these unstructured micelles at large volume fraction
frustrates and disrupts the formation of a high density colloidal ordered phase.

Above $\alpha\sim 1.3$, the system assembles into low-density micellar crystals (see Fig.~\ref{fig:snap1}C,
with the colloidal particles freely diffusing at their surfaces.  
This phase is analogous to that observed using a 
coarse-grained, soft-sphere model for the polymers~\cite{Capone}. 

{We argue that the dominant contribution to the system pressure in the  micellar regime 
comes from the free energy penalty associated with chain confinement within each micelle.
The free energy cost per micelle associated with it can be readily evaluated by  simple scaling arguments~\cite{deGennes}}, 
\begin{equation}
\Delta f\propto n \left( \frac{R_{\text g}}{R_{\text m}}\right )^2 \,,
\end{equation}
where $R_{\text m}$ is the radius of the micelle, from which we estimate
that the pressure of the system should scale with micellar radius as 
\begin{equation}
P\propto N_{\text m} n \frac{R_{\text g}^2}{R_{\text m}^5} \,,
\label{pressure}
\end{equation}
where $N_{\text m}$ is the number of micelles forming the crystal.

\begin{figure}
  \centering \includegraphics[width=\picturewidth]{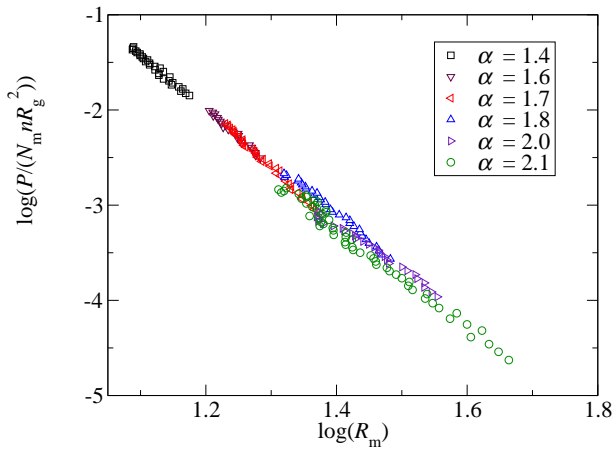}
  \caption{(Color online) Data collapse of the rescaled pressure $P/(N_{\text m} n R_{\text g}^2)$ as a function of micellar
 size $R_{\text m}$    for different values of $\alpha$. }
  \label{fig:Rm}
\end{figure}

Figure~\ref{fig:Rm} shows how all data collected for different 
combinations of colloidal radii and polymer lengths in the micellar crystal 
phase can  indeed be collapsed into the same master curve {when properly normalized}. 
A power law fit to the data, {i.e. the pressure $P$ imposed in our simulations and the corresponding measured average micellar radius}, 
yields a pressure dependance on $R_{\text{m}}$, $P\propto R_{\text{m}}^{5.4(2)}$,
 which is consistent with the eq.~\ref{pressure} for large values of $R_{\text m}$. 
Clearly, our theory  breaks down at very large densities, i.e.
small micellar radii, where long tethers
begin to radiate out of the micellar cores. This happens when the main mechanism of micellar shrinkage 
involves exclusively  colloidal expulsion from the micellar surface, causing significant thickening 
and layering of colloids in the inter-micellar regions.

{Unfortunately extending the validity of our scaling behavior to large size ratios would require performing simulations 
with very long polymers, and this becomes quickly untreatable. Nevertheless, the fact that all the data gathered 
in our simulations for different size ratios collapse onto the same master curve is strongly
 suggestive that the dominant contribution to the system pressure comes indeed from the free energy penalty associated with chain confinement.}

\subsection*{Self-avoiding chains}
\begin{figure*}
    \includegraphics[width=2.0\picturewidth]{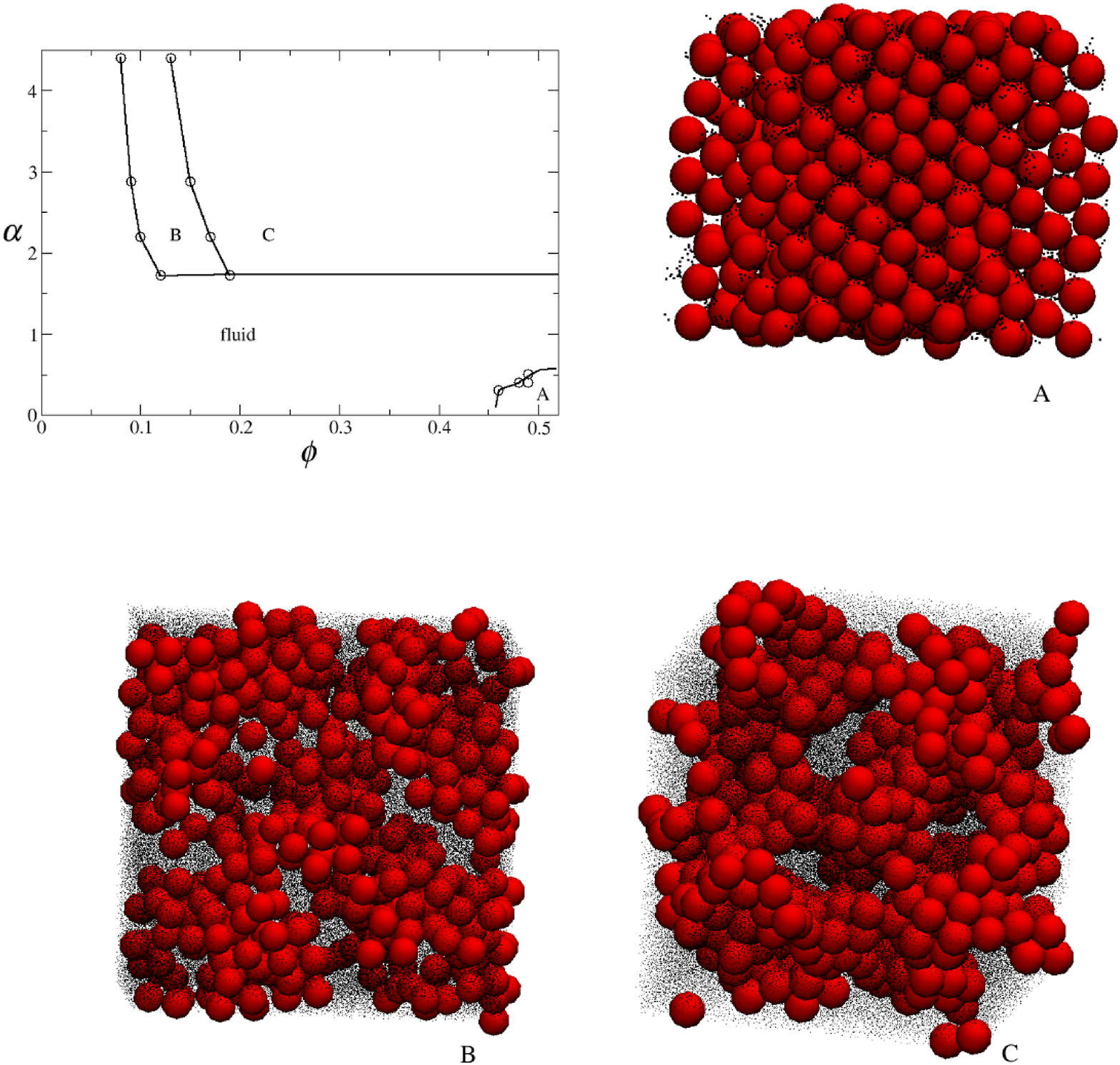}
  \caption{(Color online) Phase behavior for the case of self-avoiding chains. 
In the graph in the top-left corner of the figure,
the vertical axis indicates the polymer-colloid size ratio $\alpha$
and the horizontal axis is the colloidal volume fraction $\phi$. Snapshots of the phases in region (A), (B) and (C), depicting 
the colloidal crystal, the colloidal cluster, and the bicontinuous phase respectively are also shown.
For the sake of clarity, in snapshots (B) and (C), the polymers are depicted using a light, low density pixel representation.
}\label{SASA}
\end{figure*}

Figure~\ref{SASA} shows the phase behavior as a function of particle volume
fraction for different values of particle-to-polymer size ratio when self-avoiding chains are
connected to the colloids, and presents a quite different landscape.
We still find that for sufficiently small $\alpha$, colloids crystallize into
an FCC crystal by fitting the chains in the colloidal interstitial  spaces  (Fig.~\ref{SASA}A).
However, chains never mix to
form vacancies, and as the length of the polymer increases, the colloidal crystal becomes frustrated and eventually 
ceases to form. Unlike the case of ideal polymers, we see no evidence of a micellar phase. 
We believe this is due to the large entropic barrier
associated with overlapping multiple confined chains.
This can be estimated by  computing the confinement free energy of a polymer of length equal to the sum of all chains in the cavity, which
would grow as $\Delta F\propto (R^{\text{eff}}_{\text g}/R_m)^{3/(3\nu-1)}$, where $R^{\text{eff}}_{\text g}$ is the radius of 
gyration of a chain of length $nN$ and $\nu\simeq 3/5$. Clearly, the free energy dependence on both the 
number of chains $n$ and size of the cage $R_m$, $\Delta F\propto n^{9/4} (R_{\text g}/R_m)^{3.75}$, 
 is much stronger than what obtained for ideal chains~\cite{jun}. As a result, as soon as chains 
become confined, any significant amount of polymer overlap is highly unfavorable.

As the polymer size increases, for $0.5\leq \alpha\leq 1.75$, the dense phase presents no
colloidal order. Chains do not mix with each other and occupy the interstitial spaces in between colloids. 
The overall shape of the chains is elongated, as this geometry is entropically more favorable than a spherical one \cite{cacciuto2,sakaue,grosberg}..

As soon as $\alpha$ becomes larger than 1.75, the micellar phase found for ideal chains is replaced by a disordered 
bicontinuous phase (Fig.~\ref{SASA}C), which allows for a more effective lateral packing of the chains.
This phase is preceded by the formation of small colloidal clusters driven 
together by a combination of depletion interactions and chain-chain repulsions (Fig.~\ref{SASA}B).
The colloid-rich region presents, in both cases,  a significant degree of crystalline order.

\begin{figure}
  \centering \includegraphics[width=\picturewidth]{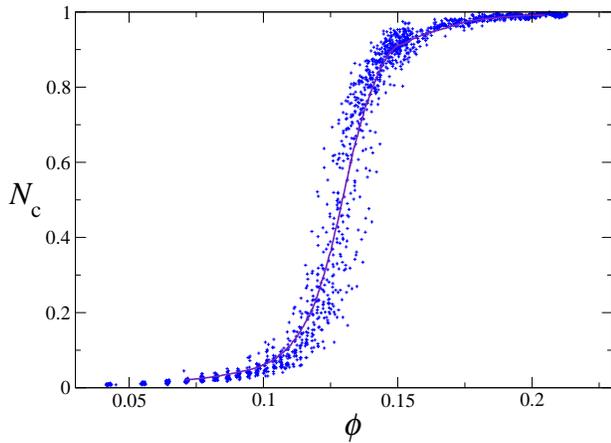}
  \caption{Size of the largest colloidal cluster, $N_c$, (normalized by the total number of colloids)
as a function of colloidal volume fraction, $\phi$, for self-avoiding tethers. 
Dots are simulations with colloids of size $\sigma_2 / \sigma_1 = 2.5$, connected to $N=75$ monomers at 
different initial configurations. The solid line is a mobile average of the data points.
}
  \label{fig:perc}\end{figure}

The cluster phase is stable within a  relatively  narrow range of volume fractions, and is promptly transformed into the bicontinuous phase as
soon as $\phi$ is sufficiently large for the clusters to merge.
Fig.~\ref{fig:perc} shows how the size of the 
largest colloidal cluster in the system, normalized 
by the total number of colloids, grows with the system volume fraction.
It is worth mentioning that colloidal clusters can grow quite thick,
and this can only be attained at the expense of the entropy of the polymers connected to the particles at 
the core of each cluster, as they need to be partially unwrapped. We believe that this free energy cost may 
actually limit the overall thickness of the clusters and incentivize linear, rather than isotropic, 
cluster growth.

The overall phenomenology in this region can again be understood in terms of chain confinement. 
It is well known \cite{Bolhuis} that the free energy cost to completely overlap two unconfined chains 
is about $2k_{\text{B}}T$, independent of the polymer length. As a result, at low volume fractions,  there isn't 
a significant driving force for self-organization. However, as $\phi$ increases and the chain sizes become smaller than 
$R_{\text g}$, $\Delta F$ acquires, as discussed above, a nontrivial dependence of the number of chains, $n$, sharing the same
volume. This leads to chain reorganization and subsequent colloidal clustering. These clusters 
 present no translational order or size monodispersity, and  are stabilized 
by their mutual effective repulsions, which extend to a surface-to-surface range that is typically smaller than $R_{\text g}$.
As soon as $\phi$ is sufficiently large, clusters merge to
further minimize chain-chain interactions and  the bicontinuous phase discussed above is formed.
For even larger volume fractions we observe significant ordering of the overall structure 
of the bicontinuous phase; however, the system sizes considered in this study are too small to
make any conclusive claim in this regard. 

The colloidal-cluster phase can be interpreted as a disordered inverted micellar phase.
We cannot a priori exclude the existence of an inverted micellar crystal phase
for even larger polymer lengths than the ones considered in this study, 
but such an analysis is out of the reach of our computational resources.

\section{Conclusions and discussion}
\label{sec:discussion}

We report the phase behavior of a system of hybrid colloids
formed by grafting a single polymer on the surface of a colloidal particle.
We find a variety of self-assembled structures as a function of polymer-colloid size ratio and volume fraction.
The structures are driven by compressing the disordered low-density states and can be understood in terms of the entropy of both tethers and colloids. 
We have identified chain confinement as the key parameter to sort out the physical mechanisms driving self-assembly in this system.

It would be interesting to test whether, for 
self-avoiding polymers,  an ordered bicontinuous phase and a crystal phase of inverted micelles
can indeed be obtained, and to study how the phase behavior presented in this
manuscript changes as a function of the number of grafted polymers.

We wish to stress that both disordered and ordered micellar phases were observed 
by Capone and collaborators \cite{Capone} while studying a system of diblock copolymers modeled as an 
ideal and self-avoiding polymer with a density-dependent effective soft-sphere potential. This seems to suggest that 
(a) the nature of the micellar phase for ideal tethers is not too sensitive to the details of the 
interaction, and (b) for $\alpha$ sufficiently large, ideal chains are indeed well-characterized by an additive
effective pair potential. The problem becomes more complicated when dealing with self-avoiding polymers. When 
multiple polymers are confined within the same region their interaction energy does not scale linearly 
with the number of chains, but as $n^{9/4}$, and up to $n^3$ for even larger densities \cite{jun}. 
This is clearly not pairwise additive. Some preliminary results obtained using an effective soft-spherical potential
to describe the polymer (to be published elsewhere) indicate  quite different phase behavior, including several 
ordered phases which are not found in our simulations with explicit polymers. This
seems to suggest that a more sophisticated coarse-graining of self-avoiding polymers is required to
obtain the correct phenomenological behavior of this system.

\section*{Acknowledgments}
This work was supported by the National Science Foundation
under CAREER Grant No. DMR-0846426.

\end{document}